\pdfoutput=1
\documentclass[cameraready]{Interspeech}

\usepackage{amsmath,amssymb}
\usepackage{booktabs}
\usepackage{multirow}
\usepackage{graphicx}
\usepackage{xcolor}
\usepackage{subcaption}
\usepackage{tikz}
\usepackage{pgfplots}
\usepackage{algorithm}
\usepackage{algpseudocode}
\usepackage{hyperref}
\usepackage{xurl}
\usepackage{microtype}
\usepgfplotslibrary{groupplots}
\pgfplotsset{compat=1.18}

\newcommand{\methodname}{DTM-Codec}
\newcommand{\dtm}{DTM}
\newcommand{\ple}{PLE}

\newcommand{\ie}{\textit{i.e.}}
\newcommand{\eg}{\textit{e.g.}}

\title{DTM-Codec: Dynamic Token Masking for VFR Speech Coding with Efficient Boundary Selection}

\author[affiliation={1}]{Hoyeol}{Sohn}
\author[affiliation={1}]{Juhan}{Nam}

\address{
    $^1$ Graduate School of Cultural Technology, KAIST, South Korea
}

\email{\{hoyso48,juhan.nam\}@kaist.ac.kr}


\begin{document}

\maketitle

\begin{abstract}
Variable frame rate (VFR) coding has recently emerged in neural speech codecs, allocating fewer frames to redundant regions and more frames to rapidly changing speech.
VFR must transmit side information about retained time steps, but prior gains are either not rigorously addressed or often minor once these overhead bits are included in total bitrate.
We present Dynamic Token Masking (DTM)-Codec, a neural speech codec that demonstrates clear gains over fixed-frame-rate baselines under a strict matched-total-bitrate protocol.
DTM keeps selected encoder tokens, fills masked positions with a learned \texttt{<MASK>} embedding, and transmits a binary keep-mask for position-aware decoding.
We further introduce Path Length Equalization (PLE), a linear-time boundary selector for VFR coding that yields well-spread adaptive segments with negligible overhead.
Across operating points, DTM-Codec broadly improves reconstruction quality and intelligibility over fixed-frame-rate baselines.
\end{abstract}

\section{Introduction}
\label{sec:intro}

Neural audio codecs have become a central tokenization layer for speech language models~\cite{borsos2023audiolm,wang2023neural,ye2025llasa}, text-to-speech synthesis~\cite{chen2025neural}, and audio generation~\cite{kreuk2023audiogen}.
The field has advanced from multi-codebook residual vector quantization (RVQ) systems such as SoundStream and EnCodec toward single-codebook tokenizers and semantic-aware designs~\cite{zeghidour2022soundstream,defossez2022high,kumar2023high,ye2024xcodec}, while recent surveys identify temporal token allocation as a key open challenge beyond codebook design~\cite{zhang2025recent,zhang2025discrete}.
Single-codebook systems such as BigCodec~\cite{xin2024bigcodec}, WavTokenizer~\cite{ji2024wavtokenizer}, and TAAE~\cite{parker2024scaling} simplify language-model integration by producing a flat token sequence, yet they operate at fixed frame rates (\eg, 25--100~Hz), assigning uniform temporal resolution regardless of information density.

Variable frame rate (VFR) coding addresses this limitation by allocating temporal resolution adaptively: fewer tokens for silence or sustained vowels, more for rapid phonetic transitions.
Several recent works explore paradigms for VFR neural codecs.
TFC-Codec~\cite{zhang2025unlocking} augments a standard codec with a plug-in module that selects among predefined temporal granularities per region.
VARSTok~\cite{zheng2025say} fine-tunes a pretrained single-codebook codec to emit variable-length tokens by clustering adjacent frames and encoding duration implicitly within the token index.
CodecSlime~\cite{ji2025codecslime} applies VFR as a post-training plugin: an optimization-based scheduler determines merge patterns on frozen encoder features, after which the decoder is adapted via two-stage fine-tuning.
FlexiCodec~\cite{huang2024flexicodec} adopts a dual-stream architecture with a frozen ASR encoder alongside a codec encoder, merging frames guided by semantic similarity to reach low frame rates.

However, whether VFR yields \emph{consistent} reconstruction gains at matched total bitrate remains an open question.
A fair comparison must account for side-information bits (position or duration) together with content bits.
VARSTok, TFC-Codec, and FlexiCodec each report bitrate-matched or reduced-rate settings, yet none provides a controlled same-architecture VFR-vs-FFR comparison in which timing-side-information overhead is explicitly included in the total bitrate~\cite{zheng2025say,zhang2025unlocking,huang2024flexicodec}.
CodecSlime includes a strict matched-rate comparison, but the reported gains are partial and concentrated on a subset of metrics~\cite{ji2025codecslime}.
In addition, optimization parity is not always controlled: VARSTok adapts a pretrained WavTokenizer, and CodecSlime applies an additional two-stage Melt-and-Cool adaptation on top of a pretrained FFR backbone to obtain its VFR variant~\cite{zheng2025say,ji2025codecslime}.
This motivates a practical question: \textit{can VFR deliver reliable reconstruction improvements at matched total bitrate when side-information overhead is explicitly accounted for?}

We address this question with \methodname{}, a 127M-parameter neural speech codec trained on LibriSpeech-960 alone~\cite{panayotov2015librispeech}---modest in both model size and data compared with recent systems that leverage large-scale multi-domain corpora and semantic teacher models~\cite{xin2024bigcodec,ye2025llasa,huang2024flexicodec}.
Despite these constraints, \methodname{} outperforms or matches larger external codecs across multiple reconstruction metrics at matched bitrate, demonstrating that well-designed VFR mechanisms can compensate for scale~\cite{xin2024bigcodec,ye2025llasa,huang2024flexicodec}.
Our contributions are as follows:
\begin{enumerate}
\item \textbf{Dynamic Token Masking (\dtm{}).} Rather than merging or pooling features, we retain selected encoder tokens and fill missing positions with a learnable \texttt{<MASK>} embedding. The binary mask is transmitted as compact position bits, enabling position-aware decoding. This formulation yields the strongest overall reconstruction among common down/up-sampling alternatives at the same bitrate.
\item \textbf{Path Length Equalization (\ple{}).} We introduce an $O(N)$ boundary selector that partitions the cumulative feature change along the encoder trajectory into equal-length segments. Compared with representative heuristic and optimization-based selectors---including those used by VARSTok, FlexiCodec, and CodecSlime---\ple{} achieves well-spread temporal coverage with negligible runtime overhead and a favorable quality--efficiency trade-off.
\item \textbf{Strict matched-total-bitrate training and evaluation.} We train and evaluate matched-rate VFR and FFR under a unified protocol that counts both content and timing-side-information bits across low-to-high frame-rate operating points. This controlled setup enables direct and fair comparisons and reveals clear and broadly VFR advantages over fixed-frame-rate baselines at low-to-mid bitrates.
\end{enumerate}
\url{https://github.com/hoyso48/DTM-Codec}.

\section{Related Work}
\label{sec:related}

\subsection{Neural Audio Codecs}
Neural audio codecs (NACs) typically follow the VQ-VAE/VQ-GAN paradigm~\cite{van2017neural}, learning an encoder--decoder with a discrete bottleneck~\cite{zeghidour2022soundstream,defossez2022high}.
SoundStream and EnCodec use residual vector quantization (RVQ) with multiple codebooks to scale bitrate~\cite{zeghidour2022soundstream,defossez2022high}.
Recent work pushes toward lower bitrates and LM-friendly token interfaces: TAAE uses a transformer autoencoder with an FSQ bottleneck and reports 400--700~bps operating points~\cite{parker2024scaling,mentzer2024finite}; BigCodec and WavTokenizer use single-codebook VQ tokenizers~\cite{xin2024bigcodec,ji2024wavtokenizer}.
Multi-resolution designs such as SNAC assign different frame rates to different quantizer layers~\cite{siuzdak2024snac}.
X-Codec~\cite{ye2024xcodec} and X-Codec 2.0~\cite{ye2025llasa} incorporate semantic feature distillation, while SpeechTokenizer~\cite{zhang2024speechtokenizer} disentangles semantic and acoustic information across RVQ layers.
DAC~\cite{kumar2023high} improves codebook utilization within the RVQGAN framework.
We build on the TAAE two-stage transformer template but use single-codebook VQ, an STFT/iSTFT front-end/back-end, and dynamic token masking for VFR coding (Section~\ref{sec:architecture}).

\subsection{Variable Frame Rate Coding in Speech Codecs}
Variable-rate discrete representation has roots in learned compression of images and audio~\cite{dieleman2021variable}; variable-rate hierarchical CPC has demonstrated acoustic unit discovery in speech~\cite{cuervo2022variable}, and recent SSL work shows that self-distillation induces syllable-level temporal structure~\cite{cho2024sdhubert}, providing linguistic motivation for adaptive-rate tokenization.
Recent VFR speech codecs mainly differ along three axes: boundary selection, timing-side-information design, and training strategy.
TFC~\cite{zhang2025unlocking} computes signal-level Shannon entropy over amplitude histograms to route each temporal window to one of three predefined granularities and fuses multi-resolution quantized tokens back to the finest timeline via granularity masks; the reported bitrate accounts for content bits only, without explicitly quantifying the mask overhead.
VARSTok~\cite{zheng2025say} applies density-peak clustering on encoder features to produce content-adaptive segments with non-uniform boundaries, and folds duration into an extended token index $\mathrm{ID} = (d{-}1) \times K + k$, enabling standard next-token prediction over an expanded vocabulary without a separate duration channel.
CodecSlime~\cite{ji2025codecslime} formulates boundary selection as a global L2-distortion minimization via dynamic programming and transmits segment lengths as explicit $\lceil\log_2 S_{\max}\rceil$ bits per frame (where $S_{\max}$ is the maximum segment span); a two-stage \emph{Melt-and-Cool} post-training recipe adapts a frozen FFR backbone to dynamic-rate operation.
FlexiCodec~\cite{huang2024flexicodec} thresholds cosine similarity between consecutive frames of a frozen ASR encoder to merge segments of up to 8 frames, transmitting 3-bit length tags; random threshold sampling during training enables continuous rate control at inference and targets a sub-10\,Hz regime.
TaDiCodec~\cite{wang2025tadicodec} takes a complementary approach: rather than adaptive selection, it operates at a fixed 6.25\,Hz and compensates for extreme compression via a text-conditioned flow-matching decoder, achieving 87.5\,bps but requiring transcription at decode time.
Compared with prior work, we focus on strict matched-total-bitrate evaluation with explicit position-bit accounting, and on a masking-based design that preserves selected token vectors exactly rather than averaging them.

\subsection{Adaptive Token Reduction for Efficient Transformers}
Adaptive token reduction in transformers is commonly grouped into pruning, pooling, and merging.
Pruning methods learn token importance and remove low-utility tokens (e.g., PoWER-BERT, LAT, Learned Token Pruning, DynamicViT, SPViT, Adaptive Token Sampling)~\cite{powerbert,lat,learnedtokenpruning,dynamicvit,spvit,ats}.
Pooling methods compress multiple tokens into learned summaries (e.g., Token Pooling, TokenLearner, Dynamic Token Pooling)~\cite{tokenpooling,tokenlearner,nawrot2023dtp}.
Merging methods combine similar tokens directly (e.g., EViT, ToMe, ToMe-SD)~\cite{evit,bolya2023tome,bolya2023tomesd}.
These methods primarily target internal compute and memory efficiency while preserving task accuracy.
Their success in transformer architectures motivated our adoption of a transformer-based codec backbone, where analogous token-level selection can be repurposed for VFR coding.
In contrast to the efficiency-oriented setting, our goal is speech coding at matched total bitrate, where the reduced token set must be transmitted together with explicit side information describing which positions were kept.

\section{Method}
\label{sec:method}

\subsection{Model Architecture}
\label{sec:architecture}

\subsubsection{Backbone}

\begin{figure*}[!t]
\centering
\includegraphics[width=\textwidth]{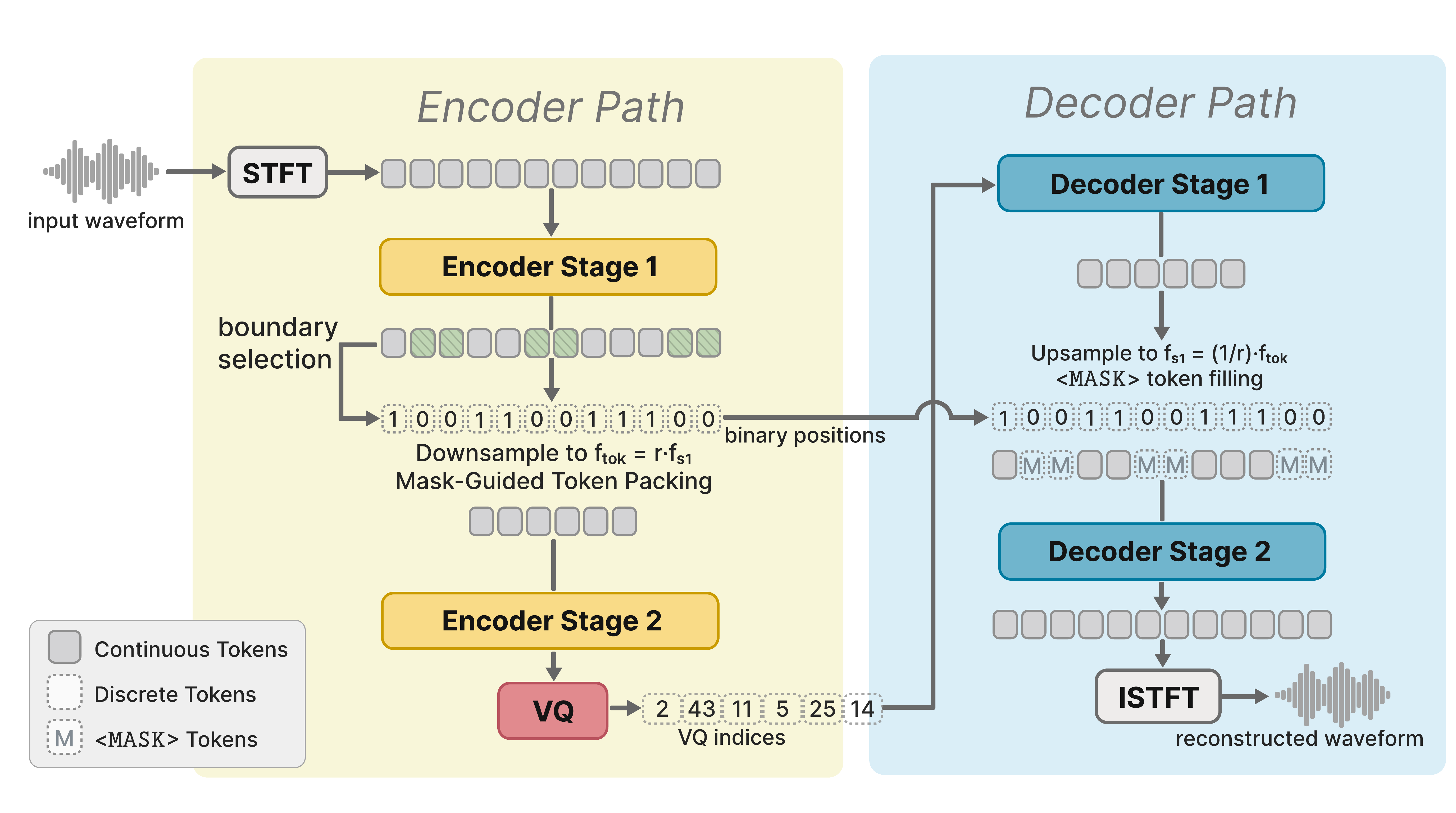}
\caption{\methodname{} architecture. \dtm{} operates between encoder stages: kept tokens ($m_t{=}1$), selected by \ple{}, are quantized via VQ and decoded; masked positions ($m_t{=}0$) are filled with a learnable \texttt{<MASK>} embedding. The binary mask $\mathbf{m}$ is transmitted as position bits.}
\label{fig:architecture}
\end{figure*}

\methodname{} builds on the two-stage transformer encoder--decoder of TAAE~\cite{parker2024scaling} (Figure~\ref{fig:architecture}).
The hierarchical encoder--decoder decomposes into a dense Stage-1 operating at the full temporal resolution and a compressed Stage-2 processing a reduced token set.
The fixed-rate strided convolution between stages can be replaced with a content-adaptive masking module without altering the rest of the pipeline, enabling controlled VFR--FFR comparisons at matched frame rates.

\subsubsection{Front-end / back-end}

TAAE uses a learned patch projection as its waveform-to-latent front-end.
We replace the front-end and back-end with an STFT/iSTFT pair, following Fourier-domain codec designs such as ComplexDec~\cite{li2025complexdec}.
For the decoder, we follow iSTFT-based reconstruction used in iSTFTNet~\cite{kaneko2022istftnet}, Vocos~\cite{siuzdakvocos}, and X-Codec 2.0~\cite{ye2025llasa}.
The decoder predicts STFT magnitude and phase, and the iSTFT synthesizes time-domain waveforms.
The STFT converts the input waveform into overlapping spectral frames at rate $f_{\mathrm{S1}}$ (\eg, 50~Hz), which are linearly projected to yield $N$ Stage-1 tokens $\mathbf{Z}^{(1)} \in \mathbb{R}^{N \times D}$.
Because the Fourier basis is fixed rather than learned, this decomposition provides stable convergence across the diverse frame rates required for VFR experiments.

\subsubsection{Quantization}

TAAE employs Finite Scalar Quantization (FSQ)~\cite{mentzer2024finite}, which distributes the bottleneck across multiple bounded scalar channels.
We replace FSQ with a single-codebook Vector Quantization (VQ) bottleneck ($|\mathcal{C}|{=}16{,}384$, $b{=}14$ bits/token) trained via straight-through estimation.
VQ assigns exactly $b = \log_2 |\mathcal{C}|$ bits per token, making bitrate accounting straightforward---an important property for the matched-total-bitrate comparisons that are central to this work.

\subsubsection{Transformer blocks}

The transformer blocks~\cite{vaswani2017attention} follow the pre-norm layout with LayerScale from TAAE~\cite{parker2024scaling}, with the following substitutions: SwiGLU~\cite{shazeer2020glu} replaces the standard FFN, RMSNorm~\cite{zhang2019rmsnorm} replaces LayerNorm, and rotary position embeddings (RoPE)~\cite{su2024roformer} provide relative position information.
$\mathrm{Enc}_1$ and $\mathrm{Dec}_2$ each use 6 transformer layers operating at the full Stage-1 resolution; $\mathrm{Enc}_2$ and $\mathrm{Dec}_1$ each use 12 layers at the compressed Stage-2 resolution.
All layers share a hidden dimension of 512 with 4 attention heads.
Following TAAE, attention uses a sliding window of 128 steps, which bounds memory cost while preserving sufficient temporal context for speech.
For VFR training, Stage-2 sequences vary in length across batch samples; we use FlashAttention's variable-length kernel~\cite{dao2022flashattention} to process packed sequences without padding.
The full model comprises approximately 127M parameters.

\subsubsection{Pipeline overview}

The complete processing pipeline is:
\begin{equation}
\label{eq:pipeline}
\begin{aligned}
\mathbf{x} &\xrightarrow{\mathrm{STFT}} \mathbf{S}
\xrightarrow{\mathrm{Enc}_1} \mathbf{Z}^{(1)}
\xrightarrow{\mathrm{DTM}} \tilde{\mathbf{Z}}^{(1)}
\xrightarrow{\mathrm{Enc}_2} \mathbf{Z}^{(2)} \\
&\xrightarrow{\mathrm{VQ}} \hat{\mathbf{Z}}^{(2)}
\xrightarrow{\mathrm{Dec}_1} \mathbf{H}
\xrightarrow{\mathrm{Up}} \hat{\mathbf{Z}}^{(1)}
\xrightarrow{\mathrm{Dec}_2} \hat{\mathbf{S}}
\xrightarrow{\mathrm{iSTFT}} \hat{\mathbf{x}}
\end{aligned}
\end{equation}

We denote the Stage-1 frame rate (determined by the STFT hop size) as $f_{\mathrm{S1}}$~(Hz) and the effective token rate after masking as $f_\text{tok} = r \, f_{\mathrm{S1}}$~(Hz), where $r$ is the keep ratio.

In TAAE, a fixed convolutional stride of~2 between $\mathrm{Enc}_1$ and $\mathrm{Enc}_2$ reduces the sequence to $N/2$ tokens~\cite{parker2024scaling}.
\methodname{} replaces this fixed downsampling with \dtm{} (Section~\ref{sec:dtm}), which retains a fraction $r$ of Stage-1 tokens---the \emph{keep ratio}---determined by a content-adaptive boundary selector.
With $r{=}0.5$, the average compression matches the original stride-2 factor, but retained positions concentrate on informative regions rather than being uniformly spaced.
The $K{=}rN$ kept tokens are processed by $\mathrm{Enc}_2$ and quantized, yielding an effective token rate $f_\text{tok}{=}r\,f_{\mathrm{S1}}$ and a content bitrate of $f_\text{tok} \times b$~bps.
A 1-bit-per-Stage-1-step binary mask is transmitted as position side information (Section~\ref{sec:bitrate}).

\subsubsection{Training objective}

Following BigCodec~\cite{xin2024bigcodec}, we train the model with an adversarial objective combined with reconstruction losses.
The discriminator ensemble comprises a multi-period discriminator (MPD)~\cite{kong2020hifigan} and a multi-scale STFT (MS-STFT) discriminator~\cite{defossez2022high}, both with least-squares GAN loss~\cite{mao2017lsgan}.
Reconstruction uses the multi-scale mel-spectrogram $L_1$ distance from DAC~\cite{kumar2023high}, computed at multiple FFT sizes and mel-bin counts, plus an $L_1$ feature-matching loss on intermediate discriminator activations~\cite{kong2020hifigan}.

\subsection{Dynamic Token Masking}
\label{sec:dtm}

\dtm{} enables variable-rate coding within the two-stage architecture.
Given the dense Stage-1 output $\mathbf{Z}^{(1)} \in \mathbb{R}^{N \times D}$, it selects a subset of positions to retain, packs them into a shorter sequence for Stage-2 processing, and transmits the binary keep-mask $\mathbf{m}$ as side information for position-aware reconstruction.

Prior VFR codecs typically downsample by averaging features within each segment and upsample by repeating the segment representative across all positions in that interval~\cite{zhang2025unlocking,ji2025codecslime,huang2024flexicodec}.
While efficient, repeat-based upsampling is brittle when boundaries are imperfect: a misaligned boundary can broadcast a single token across mismatched regions (\eg, spanning a phonetic transition), degrading reconstruction.
This motivates two design choices: mask-guided packing avoids aggregation at kept positions, and \texttt{<MASK>}-token filling treats masked slots as explicit unknowns rather than duplicating neighboring content.

A boundary selector (\eg, \ple{}, Section~\ref{sec:ple}) produces the binary mask $\mathbf{m} \in \{0,1\}^N$, where $m_t{=}1$ denotes a kept position and $m_t{=}0$ a masked one; the first token is always kept ($m_1{=}1$).
The selector is external to the codec and can be swapped at inference time (Section~\ref{sec:exp_setup}).

The kept tokens are packed into $\tilde{\mathbf{Z}}^{(1)} \in \mathbb{R}^{K \times D}$ with $K{=}\sum_t m_t$:
\begin{equation}
\tilde{\mathbf{Z}}^{(1)} = \operatorname{Pack}(\mathbf{Z}^{(1)}, \mathbf{m}) = [\mathbf{z}^{(1)}_t \mid m_t = 1].
\end{equation}
Unlike token merging~\cite{bolya2023tome} or averaging-based pooling, packing preserves the original feature vectors exactly, avoiding information loss from aggregation.
The packed sequence is then processed by the Stage-2 encoder and quantized.

After Stage-1 decoding, the output tokens $\mathbf{H} \in \mathbb{R}^{K \times D}$ are restored to the original $N$-length timeline by filling masked positions with a learnable \texttt{<MASK>} embedding $\mathbf{e}_\text{mask} \in \mathbb{R}^D$:
\begin{equation}
\hat{\mathbf{z}}^{(1)}_t =
\begin{cases}
\mathbf{h}_{\pi(t)}, & m_t = 1,\\
\mathbf{e}_\text{mask}, & m_t = 0,
\end{cases}
\quad \pi(t)=\sum_{j=1}^{t} m_j.
\end{equation}
$\mathrm{Dec}_2$ then reconstructs the full spectral representation from this mixed sequence, leveraging surrounding context to infer content at masked positions.

The mask $\mathbf{m}$ costs one bit per Stage-1 time step---the key distinction from fixed-frame-rate coding.
Because the decoder knows \emph{which} positions carry quantized content and which must be inferred, it can treat the two types differently, making the side-information overhead worthwhile (Section~\ref{sec:exp_ablation}).

\subsection{Path Length Equalization}
\label{sec:ple}

\ple{} determines which Stage-1 positions to keep (\ie, the mask $\mathbf{m}$ in Section~\ref{sec:dtm}).
It treats the encoder feature trajectory as a path in representation space and places boundaries so that each segment covers an equal amount of cumulative feature change.
This yields content-adaptive allocation---more boundaries in rapidly changing regions (\eg, phonetic transitions), fewer in stationary regions (\eg, silence)---with guaranteed well-spread coverage in a single $O(N)$ pass.

\begin{algorithm}[t]
\caption{Path Length Equalization (\ple{})}
\label{alg:ple}
\begin{algorithmic}[1]
\Require Stage-1 features $\mathbf{Z}^{(1)}=[\mathbf{z}_1,\dots,\mathbf{z}_N]$, threshold $\tau>0$
\Ensure Keep-mask $\mathbf{m}\in\{0,1\}^N$
\State $m_1 \gets 1$; $S \gets 0$; $k \gets 1$; $u \gets k\cdot\tau$
\For{$t=2$ to $N$}
    \State $d_t \gets 1 - \frac{\mathbf{z}_t^\top \mathbf{z}_{t-1}}{\lVert \mathbf{z}_t \rVert \, \lVert \mathbf{z}_{t-1} \rVert}$
    \State $S \gets S + d_t$
    \If{$S \ge u$}
        \State $m_t \gets 1$; $k \gets k+1$; $u \gets k\cdot\tau$
    \Else
        \State $m_t \gets 0$
    \EndIf
\EndFor
\State \Return $\mathbf{m}$
\end{algorithmic}
\end{algorithm}

Algorithm~\ref{alg:ple} details the single-pass procedure.
Each step distance $d_t$ measures the cosine distance between consecutive Stage-1 features; these are accumulated into a monotonic path $S$, and a boundary is placed whenever $S$ crosses the next multiple of $\tau$.
The resulting number of kept tokens is $K = 1 + \lfloor S_N/\tau \rfloor$, giving a keep ratio $r = K/N$ that depends on the utterance's total path length and the threshold.

Because the total path length $S_N = \sum_{t=2}^{N} d_t$ varies across utterances, a fixed $\tau$ yields different keep ratios per sample.
During training, $\tau$ is adjusted via a Robbins--Monro controller to match a target keep ratio $r_\text{target}$ (set to 0.5 in all experiments):
\begin{equation}
\tau \leftarrow \mathrm{clip}\bigl(\tau \cdot \exp\bigl(-\eta_t\,(\bar{r}_\text{ema} - r_\text{target})\bigr),\,
\tau_\text{min},\, \tau_\text{max}\bigr)
\end{equation}
where $\eta_t = \eta_0 / \sqrt{1 + t/T}$ is a decaying step size, $\bar{r}_\text{ema}$ is an exponential moving average of the observed keep ratio, and the update is applied every $U_\text{ctrl}$ steps.
During inference, the controller is frozen by default: the model uses the converged $\tau$ from training, yielding a keep ratio that naturally adapts per utterance based on content complexity.
Alternatively, $\tau$ can be overridden at inference to target a different operating point---a $(f_\text{tok}, B_\text{total})$ pair---without retraining.

\subsection{Bitrate Accounting}
\label{sec:bitrate}

A fair VFR evaluation must account for all transmitted information~\cite{zhang2025unlocking,zheng2025say,ji2025codecslime}.
We use ``total bitrate'' to mean the average transmitted bitrate in bps, including both content bits and VFR timing side information.
Thus, for DTM-Codec, total bitrate decomposes into content bits and position bits:
\begin{equation}
\label{eq:bitrate}
B_\text{total} = \underbrace{f_\text{tok} \cdot b}_{\text{content}} + \underbrace{p}_{\text{position}}
\end{equation}
where $f_\text{tok}$ is the effective token rate after masking, $b{=}14$ bits/token ($|\mathcal{C}|{=}16{,}384$), and $p$ is the position bitrate.
Prior VFR codecs use different timing-bit conventions. CodecSlime uses an explicit duration code where each merged token carries $\lceil \log_2 S_{\max} \rceil$ duration bits~\cite{ji2025codecslime}, yielding $p_{\text{dur}} \approx f_\text{tok}\,\lceil \log_2 S_{\max} \rceil$. VARSTok reports bitrate as $f_\text{tok}\log_2(|\mathcal{C}|\,S_{\max})$~\cite{zheng2025say}, which can be decomposed into content bits $f_\text{tok}\log_2|\mathcal{C}|$ plus an implicit duration term $p_{\text{dur}}=f_\text{tok}\log_2 S_{\max}$.

In contrast, our masking formulation transmits a binary keep-mask at Stage-1 resolution, so $p$ equals the Stage-1 frame rate $\times$ 1 bit (\eg, $p{=}100$~bps at 100~Hz), independent of local span lengths. This matches our default setting, which imposes no hard maximum span. Because duration-coded overhead depends on $S_{\max}$, the relative cost of masking versus duration coding varies with the span bound: the duration term $r f_{\mathrm{S1}}\log_2 S_{\max}$ can be smaller or larger than our fixed mask overhead $f_{\mathrm{S1}}$~\cite{zheng2025say,ji2025codecslime}.

For FFR baselines, $p{=}0$; to match total bitrate, FFR models use a larger codebook ($|\mathcal{C}|{=}65{,}536$, $b{=}16$) at the same token rate.
All comparisons are at matched $B_\text{total}$, with content, position, and total bitrate reported separately.

\begin{table*}[!t]
\caption{Reconstruction results on LibriSpeech test-clean (2,620 utterances) at matched total bitrate. ``Pos'' denotes position bits (bps). Higher is better except WER $\downarrow$; best and second-best in each bitrate block are shown in bold and \underline{underline}.}
\label{tab:main}
\centering
\footnotesize
\setlength{\tabcolsep}{3pt}
\resizebox{\textwidth}{!}{%
\begin{tabular}{lcc|ccc|cccccc}
\toprule
Model & Params & Frame Rate & Content & Pos & Total bps & UTMOSv2 & UTMOS & PESQ & STOI & Spk-Sim & WER$\downarrow$ \\
\midrule
Ground Truth & --- & --- & --- & --- & --- & 3.23 & 4.09 & 4.64 & 1.00 & 1.00 & 2.08 \\
\midrule
\multicolumn{12}{c}{\colorbox{cyan!6}{\strut\footnotesize\textit{Total bitrate $>$ 1.0 kbps}}} \\
\addlinespace[1pt]
DAC (16k)~\cite{kumar2023high} & 76M & 50 & 8000 & 0 & 8000 & 3.09 & 4.02 & \textbf{3.97} & \textbf{0.97} & \textbf{0.95} & \textbf{2.14} \\
BigCodec~\cite{xin2024bigcodec} & 159M & 80 & 1040 & 0 & 1040 & \underline{3.36} & 4.11 & 2.68 & 0.94 & 0.84 & 2.87 \\
FlexiCodec ($\tau=1.0$)~\cite{huang2024flexicodec} & 450M & 12.46 & 1234 & 37 & 1271 & 3.20 & \underline{4.20} & 2.82 & 0.94 & 0.85 & \underline{2.25} \\
DTM-Codec@80Hz & 127M & 80 & 1120 & 160 & 1280 & \textbf{3.42} & \textbf{4.20} & \underline{2.95} & \underline{0.95} & \underline{0.87} & 2.98 \\
\midrule
\multicolumn{12}{c}{\colorbox{cyan!6}{\strut\footnotesize\textit{0.8 kbps $\le$ Total bitrate $\le$ 1.0 kbps}}} \\
\addlinespace[1pt]
SNAC (12+23+47 Hz)~\cite{siuzdak2024snac} & 19.8M & 12+23+47 & 980 & 0 & 980 & 2.84 & 3.05 & 1.91 & 0.88 & 0.58 & 4.39 \\
WavTokenizer (70Hz)~\cite{ji2024wavtokenizer} & 80.9M & 75 & 900 & 0 & 900 & 2.76 & 4.00 & 2.38 & 0.91 & 0.68 & 4.32 \\
X-Codec 2.0~\cite{ye2025llasa} & 210M & 50 & 800 & 0 & 800 & \underline{3.23} & 4.13 & 2.44 & \underline{0.92} & \textbf{0.82} & \underline{2.57} \\
FlexiCodec ($\tau=0.91$)~\cite{huang2024flexicodec} & 450M & 8.26 & 818 & 25 & 843 & 3.17 & \underline{4.19} & \underline{2.46} & \underline{0.92} & 0.78 & \textbf{2.35} \\
DTM-Codec@50Hz & 127M & 50 & 700 & 100 & 800 & \textbf{3.39} & \textbf{4.22} & \textbf{2.66} & \textbf{0.93} & \underline{0.78} & 2.91 \\
\midrule
\multicolumn{12}{c}{\colorbox{cyan!6}{\strut\footnotesize\textit{0.5 kbps $\le$ Total bitrate $<$ 0.8 kbps}}} \\
\addlinespace[1pt]
TAAE (700 bps)~\cite{parker2024scaling} & 950M & 25 & 700 & 0 & 700 & 3.10 & 3.92 & 2.16 & \underline{0.91} & 0.57 & 6.21 \\
FlexiCodec ($\tau=0.867$)~\cite{huang2024flexicodec} & 450M & 6.23 & 617 & 19 & 636 & \underline{3.14} & \underline{4.14} & \underline{2.19} & 0.90 & \underline{0.71} & \textbf{2.80} \\
DTM-Codec@40Hz & 127M & 40 & 560 & 80 & 640 & \textbf{3.43} & \textbf{4.19} & \textbf{2.49} & \textbf{0.92} & \textbf{0.74} & \underline{3.27} \\
\midrule
\multicolumn{12}{c}{\colorbox{cyan!6}{\strut\footnotesize\textit{Total bitrate $<$ 0.5 kbps}}} \\
\addlinespace[1pt]
WavTokenizer (40Hz)~\cite{ji2024wavtokenizer} & 80.9M & 40 & 480 & 0 & 480 & \underline{3.11} & 3.78 & 1.88 & 0.87 & \underline{0.57} & \underline{8.16} \\
TAAE (400 bps)~\cite{parker2024scaling} & 950M & 25 & 400 & 0 & 400 & 3.03 & \underline{3.81} & \underline{2.00} & \underline{0.89} & 0.53 & 9.39 \\
VARSTok ($\tau=0.8$)~\cite{zheng2025say} & 80.9M & 34.52 & 414 & 69 & 483 & 3.08 & 3.74 & 1.69 & 0.86 & 0.50 & 10.51 \\
VARSTok ($\tau=0.7$)~\cite{zheng2025say} & 80.9M & 29.03 & 348 & 58 & 406 & 3.03 & 3.64 & 1.54 & 0.84 & 0.43 & 15.39 \\
VARSTok ($\tau=0.6$)~\cite{zheng2025say} & 80.9M & 25.02 & 300 & 50 & 350 & 3.01 & 3.58 & 1.46 & 0.82 & 0.37 & 20.69 \\
DTM-Codec@25Hz & 127M & 25 & 350 & 50 & 400 & \textbf{3.37} & \textbf{4.11} & \textbf{2.07} & \textbf{0.90} & \textbf{0.58} & \textbf{4.73} \\
\bottomrule
\end{tabular}
}
\end{table*}

\begin{table*}[!t]
\caption{Matched-rate FFR$\rightarrow$VFR comparison at the same total bitrate. Each cell reports absolute values and relative change in parentheses. In each metric column, bold and \underline{underline} on percentages mark best and second-best gains across available rates.}
\label{tab:ffr_ablation}
\centering
\footnotesize
\setlength{\tabcolsep}{4pt}
\resizebox{\textwidth}{!}{%
\begin{tabular}{cc|ccccc}
\toprule
Rate & Total bps & UTMOS (FFR$\rightarrow$VFR) & PESQ (FFR$\rightarrow$VFR) & STOI (FFR$\rightarrow$VFR) & Spk-Sim (FFR$\rightarrow$VFR) & WER (FFR$\rightarrow$VFR) \\
\midrule
25Hz   & 400  & 4.01$\rightarrow$4.11 (\underline{+2.3\%}) & 1.97$\rightarrow$2.07 (+5.1\%) & 0.89$\rightarrow$0.90 (\underline{+0.9\%}) & 0.55$\rightarrow$0.58 (+4.3\%) & 5.61$\rightarrow$4.73 (\textbf{+15.6\%}) \\
40Hz   & 640  & 4.11$\rightarrow$4.19 (+1.9\%) & 2.32$\rightarrow$2.49 (\underline{+7.0\%}) & 0.92$\rightarrow$0.92 (+0.9\%) & 0.67$\rightarrow$0.74 (\underline{+9.7\%}) & 3.77$\rightarrow$3.27 (\underline{+13.3\%}) \\
50Hz   & 800  & 4.12$\rightarrow$4.22 (\textbf{+2.4\%}) & 2.46$\rightarrow$2.66 (\textbf{+8.2\%}) & 0.92$\rightarrow$0.93 (\textbf{+1.0\%}) & 0.71$\rightarrow$0.78 (\textbf{+10.5\%}) & 3.31$\rightarrow$2.91 (+12.1\%) \\
80Hz   & 1280 & 4.19$\rightarrow$4.20 (+0.3\%) & 2.92$\rightarrow$2.95 (+1.2\%) & 0.95$\rightarrow$0.95 (-0.1\%) & 0.84$\rightarrow$0.87 (+3.4\%) & 2.54$\rightarrow$2.98 (-17.4\%) \\
\bottomrule
\end{tabular}
}
\end{table*}

\section{Experiments}
\label{sec:experiments}

\subsection{Experimental Setup}
\label{sec:exp_setup}

\noindent\textbf{Dataset and training.}\quad
All models are trained on LibriSpeech-960~\cite{panayotov2015librispeech} at 16~kHz for 600k steps on 2$\times$RTX~4090 with batch size 64, AdamW ($(\beta_1,\beta_2){=}(0.8,0.9)$), and bf16 precision.

\noindent\textbf{Model configurations.}\quad
We compare two primary model variants:
\begin{itemize}
    \item \textbf{VFR} (\ple{}): Trained from scratch with fixed keep ratio $r{=}0.5$. Single-codebook VQ ($|\mathcal{C}|{=}16{,}384$, $b{=}14$), plus 1-bit-per-step position bits.
    \item \textbf{FFR}: Trained from scratch at a fixed frame rate using uniform stride masking (every $\lfloor 1/r \rfloor$-th token kept) with no position bits. Single-codebook VQ ($|\mathcal{C}|{=}65{,}536$, $b{=}16$) to match total bitrate.
\end{itemize}

\noindent\textbf{Evaluation metrics.}\quad
Reconstruction quality is assessed with UTMOSv2~\cite{baba2024utmosv2}, UTMOS~\cite{saeki2022utmos}, PESQ (wideband)~\cite{rix2001pesq}, STOI~\cite{taal2011stoi}, speaker similarity (Spk-Sim; cosine similarity from a fine-tuned WavLM-Large speaker verification model~\cite{chen2022wavlm,baevski2020wav2vec}), and word error rate (WER) from a HuBERT-Large ASR model~\cite{hsu2021hubert}.
All external codecs are re-evaluated under the same metric pipeline using official checkpoints.

\noindent\textbf{Bitrate accounting.}\quad
For DTM-Codec, $\tau$ is set to match predefined bitrate anchors via token-rate calibration, and bps is reported from realized token rates.
For external VFR codecs (FlexiCodec and VARSTok), token rates are likewise recomputed with the same realized-rate accounting~\cite{huang2024flexicodec,zheng2025say}.

\begin{figure*}[!t]
\centering
\includegraphics[width=\textwidth]{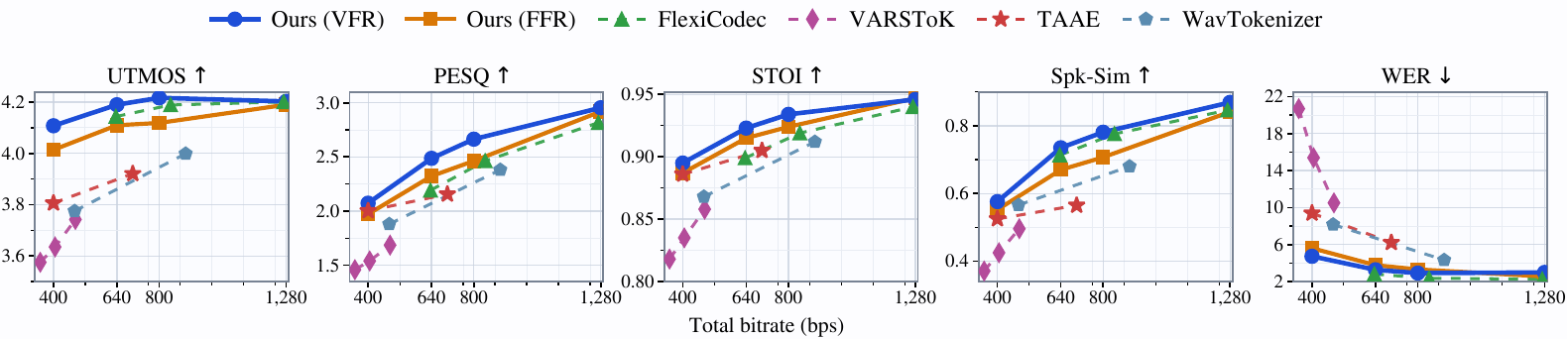}
\caption{Matched-rate comparison on LibriSpeech test-clean. Rate--distortion curves for UTMOS, PESQ, STOI, Spk-Sim, and WER compare our DTM-Codec variants (VFR and FFR) against external codecs. The x-axis reports total bitrate (bps).}
\label{fig:rd_curves}
\end{figure*}

\subsection{Main Reconstruction Results}
\label{sec:exp_main}

Table~\ref{tab:main} presents reconstruction results at four bitrate anchors (400, 640, 800, 1280~bps), corresponding to target token rates of 25, 40, 50, and 80~Hz with keep ratio $r{=}0.5$.
Each anchor yields a different content--position bitrate split: for instance, the 50~Hz anchor produces $700$~content~$+$~$100$~position~$=$~$800$~bps.
Matched-rate FFR ablations are reported separately in Table~\ref{tab:ffr_ablation};
Figure~\ref{fig:rd_curves} visualizes the rate--distortion trade-off.

Among external codecs, FlexiCodec~\cite{huang2024flexicodec} and VARSTok~\cite{zheng2025say} are VFR systems.
FlexiCodec uses an ASR-guided dual-stream RVQ architecture with dynamic frame rates (3--12.5~Hz); we evaluate three threshold settings spanning 636--1271~bps.
VARSTok applies temporal-aware density-peak clustering on WavTokenizer features at 25--35~Hz average rate.
Note that $\tau$ in FlexiCodec and VARSTok denotes their respective similarity thresholds, distinct from the PLE path-length threshold $\tau$ in Section~\ref{sec:ple}.
The remaining baselines---BigCodec~\cite{xin2024bigcodec}, X-Codec 2.0~\cite{ye2025llasa}, WavTokenizer~\cite{ji2024wavtokenizer}, DAC~\cite{kumar2023high}, SNAC~\cite{siuzdak2024snac}, and TAAE~\cite{parker2024scaling}---are FFR codecs at fixed token rates.
For all external VFR codecs, effective token rate is recomputed on LibriSpeech test-clean as total tokens divided by total audio duration over all 2{,}620 utterances.

DTM-Codec quality scales consistently with bitrate: PESQ improves from 2.07 at 400~bps to 2.95 at 1280~bps, and WER decreases from 4.73 to 2.98.
All DTM-Codec entries are trained from scratch without external semantic features.

VFR consistently outperforms FFR at matched bitrate across all metrics at 400--800~bps (Table~\ref{tab:ffr_ablation}).
The largest gains appear at 800~bps: PESQ improves by $+8.2\%$, speaker similarity by $+10.5\%$, and WER decreases by $12.1\%$ relative.
At lower bitrates the advantage is similarly clear, with PESQ improved by $7.0\%$ at 640~bps and WER reduced by $15.6\%$ at 400~bps.

At 1280~bps (80~Hz anchor), VFR retains UTMOS and PESQ advantages but FFR achieves slightly lower WER.
At this rate, consecutive Stage-1 frames already exhibit high local similarity, leaving limited room for boundary selection to remove redundant tokens, and the position-bit overhead ($160$~bps) provides diminishing returns.
This motivates operating DTM-Codec at 25--50~Hz targets.

\noindent\textbf{Subjective evaluation.}\quad
We additionally conduct a web-based, MUSHRA-inspired listening test.
19 participants each rated 8 randomly sampled utterances from LibriSpeech test-clean.
The anchor was a 3.5\,kHz low-pass filtered reference.

\begin{table}[!htbp]
\caption{MUSHRA-inspired subjective evaluation on 8 utterances from LibriSpeech test-clean.}
\label{tab:mushra}
\centering
\footnotesize
\setlength{\tabcolsep}{5pt}
\begin{tabular*}{0.99\columnwidth}{@{\extracolsep{\fill}}lccc@{}}
\toprule
System & Setting & bps & MUSHRA \\
\midrule
Hidden reference & --- & --- & 84.82 $\pm$ 2.42 \\
Anchor (3.5\,kHz LPF) & --- & --- & 51.11 $\pm$ 3.41 \\
DTM-Codec@50Hz & VFR & 800 & \textbf{81.62 $\pm$ 2.52} \\
DTM-Codec@50Hz & FFR & 800 & 78.79 $\pm$ 2.81 \\
DTM-Codec@25Hz & VFR & 400 & 71.26 $\pm$ 3.02 \\
DTM-Codec@25Hz & FFR & 400 & 68.13 $\pm$ 3.35 \\
FlexiCodec~\cite{huang2024flexicodec} & $\tau=0.91$ & 843 & 69.47 $\pm$ 3.52 \\
VARSTok~\cite{zheng2025say} & $\tau=0.7$ & 406 & 50.66 $\pm$ 3.91 \\
\bottomrule
\end{tabular*}
\end{table}

Table~\ref{tab:mushra} reports mean MUSHRA scores.
DTM-Codec VFR@50Hz (81.62) receives a higher mean score than the matched FFR baseline (78.79).
At 25\,Hz, VFR (71.26) again exceeds FFR (68.13).
FlexiCodec (69.47) and VARSTok (50.66) both score below DTM-Codec VFR.

\subsection{Ablation and Analysis}
\label{sec:exp_ablation}

\noindent\textbf{Effectiveness of masking.}\quad
Table~\ref{tab:masking} ablates the downsampler and upsampler to validate the \dtm{} design.
We test four VFR combinations (averaging vs.\ mask-guided packing $\times$ token repetition vs.\ \texttt{<MASK>} filling) under \ple{} boundaries at matched total bitrate, plus two FFR controls: mask-guided with \texttt{<MASK>} fill under uniform stride masking (FixedPattern), and convolutional down/up resampling.

\begin{table}[!t]
\caption{Downsampler $\times$ upsampler ablation at matched total bitrate, separated into VFR and FFR settings. Bold = best. Higher is better except WER.}
\label{tab:masking}
\centering
\small
\setlength{\tabcolsep}{2.5pt}
\resizebox{\columnwidth}{!}{%
\begin{tabular}{cc|cccc}
\toprule
Downsampler & Upsampler & UTMOS & PESQ & STOI & WER$\downarrow$ \\
\midrule
\multicolumn{6}{l}{\textit{VFR (PLE, $r{=}0.5$)}} \\
\midrule
Averaging   & Token repeat & 4.20 & 2.64 & 0.93 & 3.20 \\
Mask-guided & Token repeat & 4.18 & 2.62 & 0.93 & 3.12 \\
Averaging   & \texttt{<MASK>} fill & 4.19 & \textbf{2.68} & 0.93 & 3.08 \\
Mask-guided & \texttt{<MASK>} fill & \textbf{4.22} & 2.66 & \textbf{0.93} & \textbf{2.91} \\
\midrule
\multicolumn{6}{l}{\textit{FFR (fixed pattern, stride=2)}} \\
\midrule
Mask-guided & \texttt{<MASK>} fill & 4.12 & \textbf{2.46} & \textbf{0.92} & \textbf{3.31} \\
Conv & Conv & \textbf{4.13} & 2.45 & 0.92 & 3.42 \\
\bottomrule
\end{tabular}
}
\end{table}

The full \dtm{} design---mask-guided packing with \texttt{<MASK>} filling---achieves the best overall reconstruction.
Mask-guided packing preserves exact features at kept positions, avoiding information loss from averaging, while the \texttt{<MASK>} embedding provides a neutral ``unknown'' signal that lets the decoder synthesize appropriate content rather than duplicating existing features.
Among FFR controls, mask-guided with \texttt{<MASK>} fill outperforms Conv (e.g., WER $3.31$ vs.\ $3.42$), but all VFR variants surpass both FFR controls on PESQ/STOI/WER.
The full VFR mask--mask setting improves over the best FFR control by $+0.10$ UTMOS, $+0.20$ PESQ, $+0.01$ STOI, and $-0.40$ absolute WER.

\noindent\textbf{Codebook utilization.}\quad
To rule out codebook utilization as an explanation for VFR gains, Table~\ref{tab:codebook} compares 800~bps configurations across codebook sizes and frame rates.
All FFR controls use VQ; one FSQ control (FSQ levels $=[5,5,5,8,8,8]$) is included to show that our utilization trend is not tied to a specific quantizer type.

\begin{table}[!t]
\caption{Analysis on codebook utilization at 800~bps.}
\label{tab:codebook}
\centering
\small
\setlength{\tabcolsep}{3pt}
\resizebox{\columnwidth}{!}{%
\begin{tabular}{llc|cccc}
\toprule
Setting & $|C|$ & Util. & UTMOS & PESQ & STOI & WER$\downarrow$ \\
\midrule
VFR@50Hz (VQ) & 16384 & 1.00 & 4.22 & 2.66 & 0.93 & 2.91 \\
\midrule
FFR@50Hz (VQ) & 65536 & 0.97 & 4.12 & 2.46 & 0.92 & 3.31 \\
FFR@57.14Hz (VQ) & 16384 & 1.00 & 4.10 & 2.38 & 0.92 & 3.43 \\
FFR@50Hz (FSQ) & 64000 & 1.00 & 4.15 & 2.49 & 0.93 & 3.28 \\
\bottomrule
\end{tabular}
}
\end{table}

All settings achieve high utilization ($0.97$--$1.00$), so the quality gap cannot be attributed to under-utilized codebooks.
VFR@50Hz and FFR@57.14Hz both reach $1.00$ utilization, yet VFR@50Hz remains clearly superior (PESQ $2.66$ vs.\ $2.38$, WER $2.91$ vs.\ $3.43$), confirming that the gain stems from VFR itself rather than codebook utilization.
Moreover, the two FFR variants suggest that increasing bitrate by raising token rate can be detrimental: at the same 800~bps, FFR@57.14Hz underperforms FFR@50Hz (PESQ $2.38$ vs.\ $2.46$, WER $3.43$ vs.\ $3.31$).
Therefore, when comparing VFR against FFR at matched bitrate, we scale FFR bitrate by increasing codebook size at a fixed token rate.

\begin{table*}[!t]
\caption{VFR selector comparison at 800~bps under random masking + selector-specific fine-tuning (50k steps, lr$=1\times10^{-5}$, 50~Hz, VQ16384), evaluated on LibriSpeech test-clean. We additionally report model-forward RTF ($\downarrow$), selector time share ($\downarrow$), NMSE ($\downarrow$), and selector complexity. For NMSE--WER, bold and \underline{underline} indicate best and second-best within each complexity anchor.}
\label{tab:vfr_algo_ft}
\centering
\footnotesize
\setlength{\tabcolsep}{4pt}
\resizebox{\textwidth}{!}{%
\begin{tabular}{lclcc|cccccc}
\toprule
Algorithm & $S_{\max}$ & Complexity & RTF$\downarrow$ & Share (\%$\downarrow$) & NMSE$\downarrow$ & UTMOS$\uparrow$ & PESQ$\uparrow$ & STOI$\uparrow$ & Spk-Sim$\uparrow$ & WER$\downarrow$ \\
\midrule
\multicolumn{11}{c}{\colorbox{cyan!6}{\strut\footnotesize\textit{$O(N)$ selectors}}} \\
Random Masking & --- & $O(N)$ & 0.0035 & 2.32 & 0.0650 & 4.180 & 2.492 & 0.923 & 0.706 & 3.431 \\
PLE & none & $O(N)$ & 0.0035 & 1.92 & \underline{0.0393} & \underline{4.201} & \textbf{2.616} & \textbf{0.929} & \textbf{0.734} & \underline{3.095} \\
PLE & 4 & $O(N)$ & 0.0035 & 2.47 & \textbf{0.0391} & \textbf{4.208} & \underline{2.602} & \underline{0.928} & \underline{0.725} & \textbf{3.060} \\
Similarity-Threshold~\cite{huang2024flexicodec} & none & $O(N)$ & 0.0035 & 0.93 & 0.0624 & 4.109 & 2.469 & 0.922 & 0.723 & 3.903 \\
Similarity-Threshold~\cite{huang2024flexicodec} & 4 & $O(N)$ & 0.0035 & 1.43 & 0.0453 & 4.191 & 2.570 & 0.927 & \underline{0.725} & 3.224 \\
\midrule
\multicolumn{11}{c}{\colorbox{cyan!6}{\strut\footnotesize\textit{$O(N^2)$ selectors}}} \\
Peak Clustering~\cite{zheng2025say} & none & $O(N^2)$ & 0.0071 & 50.08 & \textbf{0.0380} & \underline{4.206} & 2.588 & 0.928 & 0.725 & 3.119 \\
Peak Clustering~\cite{zheng2025say} & 4 & $O(N^2)$ & 0.0072 & 50.09 & \underline{0.0388} & \textbf{4.207} & 2.583 & \underline{0.927} & 0.724 & 3.131 \\
Greedy Merging & none & $O(N^2)$ & 0.0098 & 64.58 & 0.0467 & 4.194 & \textbf{2.603} & \textbf{0.928} & \textbf{0.733} & \underline{3.096} \\
Greedy Merging & 4 & $O(N^2)$ & 0.0108 & 67.33 & 0.0436 & 4.199 & \underline{2.600} & \underline{0.928} & \underline{0.727} & \textbf{3.060} \\
\midrule
\multicolumn{11}{c}{\colorbox{cyan!6}{\strut\footnotesize\textit{$O(NKS_{\max})$ selectors}}} \\
DP Segmentation~\cite{ji2025codecslime} & 4 & $O(NKS_{\max})$ & 0.0248 & 85.85 & 0.0333 & 4.204 & 2.628 & 0.929 & 0.731 & 2.954 \\
\bottomrule
\end{tabular}
}
\end{table*}

\subsection{VFR Algorithm Comparison}
\label{sec:exp_ple}

Table~\ref{tab:vfr_algo_ft} compares \ple{} with representative VFR boundary selectors under selector-specific fine-tuning at 800~bps (50~Hz, VQ16384).
Following CodecSlime~\cite{ji2025codecslime}, which trains a VFR foundation model via random downsampling and applies structured selection at inference, we adopt a related protocol: each position is masked as $m_t \sim \mathrm{Bernoulli}(r)$ during base training, and each selector is then fine-tuned from the same random-masking checkpoint for 50k steps (lr$=1\times10^{-5}$).

All selectors share the same checkpoint, input features $\mathbf{Z}^{(1)}=[\mathbf{z}_1,\dots,\mathbf{z}_N]$, and output mask $\mathbf{m}\in\{0,1\}^N$, so differences reflect selector behavior alone.
Let $\mathrm{sim}(t) = \mathbf{z}_t^\top \mathbf{z}_{t+1} / (\lVert\mathbf{z}_t\rVert\lVert\mathbf{z}_{t+1}\rVert)$ denote adjacent cosine similarity, $N$ the sequence length, $K$ the number of kept tokens, and $S_{\max}$ the maximum segment span.

In addition to reconstruction metrics, Table~\ref{tab:vfr_algo_ft} reports three diagnostics.
RTF is model-forward real-time factor (lower is better).
Share is the fraction of forward time spent in boundary selection (\%).
NMSE is merge-space normalized mean-squared error,
\(\mathrm{NMSE}=\sum_t \lVert \mathbf{z}_t-\hat{\mathbf{z}}_t \rVert^2 / (\sum_t \lVert \mathbf{z}_t \rVert^2+\epsilon)\), where $\hat{\mathbf{z}}_t$ is the segment mean containing $t$; lower indicates better feature preservation.

\noindent\textbf{Similarity-Threshold Merging}~\cite{huang2024flexicodec} merges contiguous runs with similarity above $\tau$ into single segments, keeping run starts as boundaries; an optional span cap $S_{\max}$ prevents overlong runs.
This matches FlexiCodec's dynamic frame-merging criterion; complexity is $O(N)$.

\noindent\textbf{Temporal-Aware Density Peak Clustering}~\cite{zheng2025say} computes pairwise similarities to obtain local density $\rho_t$, peak distance $\delta_t$, and seed score $s_t=\rho_t\delta_t$.
Clusters expand bidirectionally from the highest seeds under the VARSTok criterion $\phi(\mathbf{x}_{i^*},\mathbf{x}_t)-\beta s_t>\tau$ with temporal contiguity and span limit $S_{\max}$; cluster starts become boundaries.
This follows Algorithm~1 in VARSTok; complexity is $O(N^2)$.

\noindent\textbf{DP Segmentation}~\cite{ji2025codecslime} partitions the sequence into $K$ contiguous segments minimizing total SSE:
$\min \sum_{i=1}^{K} \sum_{t \in \mathcal{S}_i} \lVert \mathbf{z}_t - \bar{\mathbf{z}}_{\mathcal{S}_i} \rVert^2$,
where $\bar{\mathbf{z}}_{\mathcal{S}_i}$ is the segment mean, solved via exact DP with recurrence $C[j,k] = \min_{s \le S_{\max}} \{C[j{-}s,\,k{-}1] + L(j,s)\}$ and $O(1)$ segment costs from prefix sums.
This matches the contiguous DP scheduler in CodecSlime; here we compare boundary computation only.
Complexity is $O(NKS_{\max})$.

\noindent\textbf{Greedy Merging} iteratively finds the pair $(t,t{+}1)$ with maximum $\mathrm{sim}(t)$ and removes $t{+}1$, repeating until $K$ tokens remain.
Complexity is $O(N^2)$.

DP achieves the best PESQ and lowest NMSE, confirming that globally optimal segmentation produces the highest-fidelity boundaries.
However, DP is about $7\times$ slower than \ple{} (RTF $0.0248$ vs.\ $0.0035$) and spends $\sim$85.9\% of forward time in selector computation, whereas \ple{} and similarity-threshold merging keep selector overhead near 1--3\%.
\ple{} and similarity-threshold merging match in speed (both RTF $\approx 0.0035$), but \ple{} maintains substantially stronger quality (e.g., WER $3.06$ vs.\ $3.22$ for $S_{\max}{=}4$), yielding a better quality--efficiency trade-off.
Uncapped similarity-threshold merging degrades sharply (WER $3.90$) because one-shot thresholding can over-merge adjacent tokens; constraining the span ($S_{\max}{=}4$) helps only partially (WER $3.22$).
\ple{} avoids this over-merging without iterative updates and remains stable under span capping (none vs.\ $S_{\max}{=}4$: PESQ $2.616$ vs.\ $2.602$, WER $3.095$ vs.\ $3.060$).
Training every selector from scratch is impractical because expensive selectors (DP, clustering, Greedy) increase per-step time by $3$--$7\times$; we therefore use a lightweight fine-tuning protocol.
Even so, from-scratch \ple{} training yields better reconstruction than \ple{} fine-tuning (PESQ $2.66$ vs.\ $2.62$, WER $2.91$ vs.\ $3.10$ at 800~bps; compare Tables~\ref{tab:main} and~\ref{tab:vfr_algo_ft}), confirming that dedicated end-to-end \ple{} training remains the strongest setting.

\begin{table*}[!t]
\caption{ARCH speech benchmark~\cite{la2024benchmarking} (post-VQ). We report classification accuracy (ACC) and macro-averaged F1 for each dataset, plus ARCH-ACC/ARCH-F1 (dataset-wise mean ACC/F1 over RAVDESS, EMOVO, AudioMNIST, and SLURP). All values are percentages. $|\mathcal{C}|$ = single-codebook size. All rows are re-evaluated under the same protocol. $^\dagger$Uses Wav2Vec2-BERT semantic features.}
\label{tab:semantic}
\centering
\small
\setlength{\tabcolsep}{6.2pt}
\begin{tabular}{llc|cc|cc|cc|cc|cc}
\toprule
\multirow{2}{*}{Model} & \multirow{2}{*}{Setting} & \multirow{2}{*}{$|\mathcal{C}|$} & \multicolumn{2}{c|}{RAVDESS} & \multicolumn{2}{c|}{EMOVO} & \multicolumn{2}{c|}{AudioMNIST} & \multicolumn{2}{c|}{SLURP} & \multicolumn{2}{c}{ARCH} \\
\cmidrule(lr){4-13}
 & & & ACC & F1 & ACC & F1 & ACC & F1 & ACC & F1 & ACC & F1 \\
\midrule
DTM-Codec@25Hz & VFR & 16{,}384 & 32.64 & 28.61 & \textbf{27.38} & \textbf{22.75} & 65.84 & 65.45 & 7.60 & \underline{1.19} & 33.37 & 29.50 \\
DTM-Codec@25Hz & FFR & 65{,}536 & 35.42 & 33.56 & 26.70 & \underline{22.13} & 65.07 & 64.77 & 7.45 & 1.06 & 33.66 & 30.38 \\
\cmidrule(lr){1-13}
DTM-Codec@40Hz & VFR & 16{,}384 & 32.99 & 28.97 & 21.09 & 17.12 & \underline{70.69} & \underline{70.45} & 7.61 & 1.15 & 33.09 & 29.42 \\
DTM-Codec@40Hz & FFR & 65{,}536 & \underline{37.50} & \underline{35.22} & 24.83 & 20.04 & 65.73 & 65.65 & 7.00 & 0.95 & 33.77 & \underline{30.46} \\
\cmidrule(lr){1-13}
DTM-Codec@50Hz & VFR & 16{,}384 & \textbf{37.85} & \textbf{36.01} & 21.94 & 14.53 & 68.26 & 68.05 & 7.11 & 0.96 & 33.79 & 29.89 \\
DTM-Codec@50Hz & FFR & 65{,}536 & 34.72 & 31.69 & 24.49 & 18.71 & 69.63 & 69.55 & 7.23 & 1.05 & \underline{34.02} & 30.25 \\
\cmidrule(lr){1-13}
DTM-Codec@80Hz & VFR & 16{,}384 & \underline{37.50} & 33.25 & 23.81 & 16.40 & 58.62 & 58.13 & 6.97 & 0.84 & 31.73 & 27.16 \\
DTM-Codec@80Hz & FFR & 65{,}536 & 36.81 & 33.93 & 26.19 & 19.56 & \textbf{70.88} & \textbf{70.76} & 7.00 & 0.96 & \textbf{35.22} & \textbf{31.30} \\
\midrule
X-Codec 2.0$^\dagger$~\cite{ye2025llasa} & 50Hz & 65{,}536 & 37.15 & 32.88 & 20.75 & 15.59 & 68.49 & 68.15 & \textbf{7.74} & \textbf{1.22} & 33.53 & 29.46 \\
BigCodec~\cite{xin2024bigcodec} & 80Hz & 8{,}192 & 36.11 & 34.43 & 17.18 & 12.38 & 65.84 & 65.74 & \underline{7.67} & 1.05 & 31.70 & 28.40 \\
VARSTok~\cite{zheng2025say} & $\tau{=}0.8$ & 4{,}096 & 27.43 & 24.01 & \underline{27.21} & 21.39 & 60.42 & 60.18 & 7.62 & 1.13 & 30.67 & 26.68 \\
VARSTok~\cite{zheng2025say} & $\tau{=}0.7$ & 4{,}096 & 27.08 & 23.74 & 25.68 & 20.27 & 61.35 & 61.09 & 7.49 & 1.07 & 30.40 & 26.54 \\
VARSTok~\cite{zheng2025say} & $\tau{=}0.6$ & 4{,}096 & 24.31 & 21.03 & 24.83 & 19.34 & 62.36 & 62.10 & 7.28 & 1.00 & 29.69 & 25.87 \\
WavTokenizer~\cite{ji2024wavtokenizer} & 75Hz & 4{,}096 & 27.43 & 22.50 & 20.41 & 15.80 & 56.62 & 56.18 & 7.12 & 0.82 & 27.90 & 23.82 \\
WavTokenizer~\cite{ji2024wavtokenizer} & 40Hz & 4{,}096 & 24.65 & 21.80 & 26.19 & 18.69 & 50.12 & 49.00 & 6.69 & 0.53 & 26.91 & 22.50 \\
\bottomrule
\end{tabular}
\end{table*}

\subsection{OOD Non-English MLS Evaluation}
\label{sec:exp_mls}

To assess out-of-domain generalization, we evaluate on the MLS non-English subset~\cite{pratap2020mls} (7 languages $\times$ 100 utterances = 700 utterances) with waveform-level metrics (Table~\ref{tab:mls_ood}).
For external baselines, we follow the operating points used in Table~\ref{tab:main}: FlexiCodec with $\tau=0.91$ and VARSTok with $\tau=0.7$.

\begin{table}[!t]
\caption{MLS non-English OOD waveform-level evaluation on 700 utterances (7 languages). Ours compares matched-rate FFR and VFR at the same bps; each VFR cell reports absolute value and relative change.}
\label{tab:mls_ood}
\centering
\footnotesize
\setlength{\tabcolsep}{2.2pt}
\resizebox{\columnwidth}{!}{%
\begin{tabular}{ll|cccc}
\toprule
Model & bps & UTMOS$\uparrow$ & PESQ$\uparrow$ & STOI$\uparrow$ & Spk-Sim$\uparrow$ \\
\midrule
\multicolumn{6}{l}{DTM-Codec} \\
@25Hz FFR & 400 & 2.97 & 1.86 & 0.87 & 0.61 \\
\shortstack[l]{@25Hz VFR\\\strut} & \shortstack[c]{400\\\strut} & \shortstack[c]{3.09\\+3.9\%} & \shortstack[c]{1.94\\+4.2\%} & \shortstack[c]{0.88\\+1.3\%} & \shortstack[c]{0.65\\+7.9\%} \\
@40Hz FFR & 640 & 3.05 & 2.19 & 0.90 & 0.72 \\
\shortstack[l]{@40Hz VFR\\\strut} & \shortstack[c]{640\\\strut} & \shortstack[c]{3.15\\+3.2\%} & \shortstack[c]{2.30\\+5.2\%} & \shortstack[c]{0.91\\+1.1\%} & \shortstack[c]{0.79\\+10.7\%} \\
@50Hz FFR & 800 & 3.04 & 2.31 & 0.91 & 0.75 \\
\shortstack[l]{@50Hz VFR\\\strut} & \shortstack[c]{800\\\strut} & \shortstack[c]{3.16\\\textbf{+4.0\%}} & \shortstack[c]{2.50\\\textbf{+8.3\%}} & \shortstack[c]{0.92\\\textbf{+1.4\%}} & \shortstack[c]{0.83\\\textbf{+11.0\%}} \\
@80Hz FFR & 1280 & 3.07 & 2.72 & 0.94 & 0.87 \\
\shortstack[l]{@80Hz VFR\\\strut} & \shortstack[c]{1280\\\strut} & \shortstack[c]{3.00\\$-$2.3\%} & \shortstack[c]{2.66\\$-$2.3\%} & \shortstack[c]{0.93\\$-$0.5\%} & \shortstack[c]{0.90\\+3.4\%} \\
\midrule
FlexiCodec~\cite{huang2024flexicodec} & 905 & 2.99 & 2.33 & 0.90 & 0.84 \\
VARSTok~\cite{zheng2025say} & 402 & 2.60 & 1.49 & 0.81 & 0.44 \\
\bottomrule
\end{tabular}
}
\end{table}

The same trend holds on OOD speech: VFR consistently improves over FFR at 400--800~bps, while maintaining competitive quality against external codecs at comparable bitrate.

\subsection{Semantic Evaluation}
\label{sec:exp_semantic}

Table~\ref{tab:semantic} evaluates semantic quality on the ARCH speech benchmark~\cite{la2024benchmarking} following VARSTok~\cite{zheng2025say}.
Four datasets (RAVDESS, EMOVO, AudioMNIST, SLURP) are evaluated via linear probing on frozen post-VQ embeddings with temporal average pooling.
A linear classifier is trained for 200 epochs (AdamW, lr$=$0.001, linear warmup + decay); the tokenizer remains frozen.

ARCH evaluates global semantic content (emotion, language, digit identity, intent) via utterance-level classification on temporally pooled embeddings~\cite{la2024benchmarking}.
DTM-Codec remains competitive without semantic supervision: at 50~Hz it reaches 37.85/36.01 (RAVDESS ACC/F1) and 68.26/68.05 (AudioMNIST ACC/F1), while X-Codec 2.0 leads on SLURP (7.74/1.22).

Across matched-rate pairs, FFR often scores slightly higher than VFR on semantic probing, especially on AudioMNIST.
This reflects the nature of the benchmark: ARCH tasks measure global attributes well captured by any token after temporal pooling, whereas VFR's advantage lies in selectively retaining phonetically informative positions---a property that benefits reconstruction (Section~\ref{sec:exp_main}) but does not necessarily improve global classification.

Among external codecs, codebook size $|\mathcal{C}|$ correlates more strongly with ACC/F1 than frame rate or VFR/FFR distinction~\cite{ye2025llasa,xin2024bigcodec,zheng2025say,ji2024wavtokenizer}.
Systems with $|\mathcal{C}|{=}65{,}536$ (X-Codec 2.0, our FFR) consistently lead, followed by $|\mathcal{C}|{=}8{,}192$--$16{,}384$ (BigCodec, our VFR), while $|\mathcal{C}|{=}4{,}096$ (VARSTok, WavTokenizer) trails.
A larger codebook encodes more information per token ($\log_2|\mathcal{C}|$ bits), directly benefiting global probing.
This pattern is partly confounded by architectural differences (e.g., X-Codec~2.0 uses a Wav2Vec2-BERT teacher), but holds across systems without semantic supervision, suggesting that codebook capacity matters more than frame-rate strategy for global semantic content.

\section{Conclusion}
\label{sec:conclusion}

We presented \methodname{}, a neural speech codec that broadly outperforms fixed-rate baselines under strict matched-total-bitrate evaluation with explicit position-bit accounting.
Dynamic Token Masking preserves selected features exactly while marking missing positions for position-aware decoding, and Path Length Equalization provides a linear-time boundary selector that rivals costlier alternatives with negligible overhead.
With only 127M parameters trained on LibriSpeech-960, \methodname{} is competitive with substantially larger codecs and generalizes to unseen languages.
Future work includes general audio, streaming, and VFR-token-based speech generation.

\section{Acknowledgements}
This work was supported by the National Research Foundation of Korea (NRF) grant funded by the Korea government (MSIT) (No. RS-2023-00222383).

\section{Use of Generative AI Disclosure}
Generative AI tools were used solely for grammar correction and manuscript formatting, not for generating scientific content.

\bibliographystyle{IEEEtran}
\bibliography{references}

\end{document}